\documentclass{article}
\usepackage{spconf,amsmath,graphicx}
\usepackage{xcolor}
\usepackage{colortbl}
\usepackage{multirow}
\usepackage{booktabs}

\usepackage[numbers,sort&compress]{natbib}
\usepackage[textsize=tiny,textwidth=0.7in, disable]{todonotes}

\title{ Towards Language-Independent Face-Voice Association with Multimodal Foundation Models}

\name{Aref Farhadipour, Teodora Vukovic, Volker Dellwo }
\address{Department of Computational Linguistics, University of Zurich 
\\aref.farhadipour@uzh.ch}

\begin{document}
%
\maketitle
\begin{abstract}
This paper describes the UZH-CL system submitted to the FAME2026 Challenge. The challenge focuses on cross-modal verification under unique multilingual conditions, specifically unseen and unheard languages. Our approach investigates two distinct architectures, consisting of a baseline dual-encoder system trained from scratch using contrastive and orthogonal projection losses, and a foundation model approach leveraging ImageBind with LoRA. To address the data scarcity and language constraints of the challenge, we curated an external Arabic dataset from VoxBlink. 
Our best-performing system, ImageBind-LoRA, demonstrates remarkable cross-lingual generalization: despite being fine-tuned exclusively on Arabic audio, it achieved an EER of 24.73\% on the evaluation set (English and German), securing 2nd place in the competition.

\end{abstract}
\section{Introduction}
\label{sec:intro}

Person recognition systems have traditionally relied on unimodal biometrics, utilizing either voice \cite{farhadipour2024analysis} or face \cite{farhadipour2023facial} analysis. However, the rapidly evolving landscape of human-machine interaction has created a high demand for multimodal systems, where the effective integration of these distinct signals is superior \cite{farhadipour2024comparative, farhadipour2025multimodal}. A critical requirement for such systems is resilience to missing data. Even when one modality becomes unexpectedly unavailable, the system should remain capable of performing cross-modal verification by relying on a robust face–voice association \cite{saeed2023single, saeed2024synopsis}. Another challenge is the impact of language variation on speaker recognition systems \cite{zhang2025quantifying}. The Face-voice Association in Multilingual Environments (FAME) 2026 Challenge \cite{moscati2025face} specifically targets this complex problem. Leveraging our extensive prior experience in the NIST SRE2024 multimodal challenge \cite{farhadipour2025cl}, UZH-CL Team participated in this benchmark and developed a language-independent framework that achieved second place on the final leaderboard.

In this work, we propose two systems for face-voice association consisting of a CLIP-based dual-encoder model and an ImageBind-based model \cite{girdhar2023imagebind}. To enhance robustness, we fuse their outputs using Z-score normalization only on the development data. The ImageBind subsystem is further adapted through Low-Rank Adaptation (LoRA) \cite{hu2022lora} to better align voice–face representations. A key component of our strategy was the curation of a clean external Arabic dataset for ImageBind fine-tuning. By training the foundation model exclusively on Arabic data and evaluating it on English, German, and Urdu, we show that the system learns robust, language-independent identity representations.

\section{Methodology}
\label{sec:method}

\subsection{System 1: Dual-Branch Gated Encoder}
Our first system adopts a CLIP-style dual-encoder architecture.
For the voice modality, we utilized a ResNet34 encoder pretrained on CN-Celeb \cite{wang2023wespeaker}. For the face modality, we employed YOLO model for face detection and then an IR101-ArcFace model trained on WebFace12M \cite{kim2022adaface}. Both branches project features into a shared 128-dimensional embedding space via two-layer fully connected networks with ReLU activation.

A gated fusion mechanism dynamically weights the contribution of voice and face embeddings before combining them. The model is optimized using a weighted combination of Symmetric Contrastive Loss, Classification Loss, and Orthogonal Projection Loss (OPL). The OPL enforces geometric constraints, encouraging intra-class compactness and inter-class separability. 

\subsection{System 2: ImageBind-LoRA }
Our superior system leverages ImageBind, a foundation model pretrained to align six modalities (including audio and vision) into a common embedding space. We utilize the Audio and Vision branches of ImageBind. The Audio branch processes Mel-spectrograms via a 12-layer ViT-B transformer. The Vision branch processes face images via a 24-layer ViT-L transformer. Both branches output 768-dimensional embeddings \cite{girdhar2023imagebind}.

To adapt this large model (approx. 1.2B parameters) to the voice-face verification task without catastrophic forgetting or high computational cost, we employ LoRA. We freeze the entire pretrained backbone and inject trainable rank decomposition matrices into the attention layers of the transformer blocks. With a LoRA rank 4, this reduces the trainable parameters to 5.1M.

The optimization process followed a specific two-stage configuration: the first stage was trained for 5 epochs with a learning rate of $10^{-3}$ and a batch size of 32 just for classifier head, while the second stage (LoRA fine-tuning) ran for 15 epochs with a reduced learning rate of $10^{-4}$ and a batch size of 16. We employ a Symmetric Contrastive Loss with Hard Negative Mining. 

\subsection{Score Fusion Strategy}
During the \textbf{development} phase, we employed Z-score normalization to fuse scores from four sources: 1) Challenge Baseline \cite{nawaz2021cross, saeed2022fusion}, 2) CLIP-Style, 3) ImageBind trained with official training data, and 4) ImageBind trained with Arabic dataset.
The final score is the arithmetic mean of the Z-scores. Because of the limited number of tokens allowed for score submission on the CodaBench website, we could not use a fusion strategy during the evaluation phase.

\section{Experimental Setup}
\label{sec:setup}

\subsection{Dataset}

We utilized the official MAV-Celeb splits provided by the challenge for training, development and evaluation.

\textbf{External Arabic Dataset:} A critical innovation in our submission was the curation of an external dataset to ensure language robustness. We constructed a subset from the VoxBlink1 and VoxBlink2 datasets \cite{lin2024voxblink2}. To ensure the data was strictly Arabic, we applied a two-step language recognition: 1) Text-based filtering to select identities labeled as Arabic speakers, and 2) Whisper-based language identification to verify the spoken language and discard non-Arabic speech segments. This dataset consists of 367,778 utterances from 6,059 Arabic speakers and contains 571 hours of face–voice YouTube data.





All models were trained on a compute node equipped with eight NVIDIA 4090 (24GB) GPUs. We utilized the AdamW optimizer coupled with a Cosine Annealing scheduler to adjust the learning rate during training. 

\section{Results}
\label{sec:results}

\subsection{Development Phase: 4-Model Fusion}
We first evaluated our subsystems and fusion strategy on the development set. Table \ref{tab:dev_results} details the performance. In all cases we trained a separate system for each one of the seen languages.

\begin{table}[h]
\centering
\caption{Development Set Performance (English/Urdu data). Comparison of individual subsystems and the Z-Score Fusion.}
\label{tab:dev_results}
\resizebox{\columnwidth}{!}{
\begin{tabular}{llcc}
\toprule
\textbf{System} & \textbf{Training Data} & \textbf{Config} & \textbf{EER (\%)} \\
\midrule
Baseline & Official & FOP Algo & 36.00 \\
CLIP-Style & Official & ResNet/ArcFace & 37.00 \\
ImageBind-Official & Eng/Urdu & LoRA ($r=2$) & 37.00 \\
ImageBind-Arabic & Arabic  & LoRA ($r=4$) & 28.00 \\
\midrule
\textbf{Z-Score Fusion} & \textbf{All 4 Models} & \textbf{-} & \textbf{26.44} \\
\bottomrule
\end{tabular}
}
\end{table}

The results highlight a significant gap between models trained on limited official data ($\sim$36-37\%) and the model trained on the larger Arabic dataset (28\%). The increased LoRA rank ($r=4$) combined with diverse data allowed the model to learn more robust identity features. The Z-score fusion further improved performance to 26.44\%, proving the complementary nature of the different architectures.

\subsection{Evaluation Phase: Final Submission}

As shown in Table \ref{tab:results}, the ImageBind-LoRA system significantly outperformed the CLIP-style baseline. The Hybrid pipeline combined ImageBind (for English/German seen/unseen) and the CLIP-style model (for English-unseen) but proved less effective than the pure ImageBind approach.

\begin{table}[h]
\centering
\caption{Performance comparison on FAME 2026 Evaluation Set (English/German data). }
\label{tab:results}
\resizebox{0.7\columnwidth}{!}{
\begin{tabular}{lcc}
\toprule
\textbf{System} & \textbf{Training Data} & \textbf{EER (\%)} \\
\midrule
CLIP-style & German + Urdu & $\sim$48.0 \\
Hybrid Pipeline & Mixed & $\sim$31.0 \\
\textbf{ImageBind-LoRA} & \textbf{Arabic} & \textbf{24.73} \\
\bottomrule
\end{tabular}
}
\end{table}

The ImageBind model, trained on Arabic, achieved consistent performance across English, German, and Urdu test sets. This confirms our hypothesis that fine-tuning a multimodal foundation model with LoRA allows it to capture identity-specific features invariant to the spoken language.

\section{Conclusion}
\label{sec:conclusion}

In the FAME 2026 Challenge, Team UZH-CL achieved 2nd place by proposing a robust, language-agnostic framework. Our experiments highlight that standard supervised learning on small, multilingual datasets (System 1) is prone to overfitting. Conversely, leveraging the ImageBind foundation model with parameter-efficient LoRA fine-tuning (System 2) enables superior generalization. Crucially, by training on a completely disjoint language (Arabic), we demonstrated that the learned face-voice associations are not language-dependent, satisfying the core objective of the challenge.


\bibliographystyle{IEEEbib}
\bibliography{refs}

\end{document}